\documentclass[runningheads]{llncs}

\usepackage{todonotes}
\usepackage[utf8]{inputenc}
\usepackage{microtype}

\usepackage{amssymb}
\usepackage{graphicx}
\usepackage[hyphens]{url}
\usepackage[inline]{enumitem}
\usepackage{balance}
\usepackage{cite}

\usepackage{eurosym}
\usepackage{enumitem}

\begin{document}
\title{Deep Learning Application in Security and Privacy -- Theory and Practice: A Position Paper}
%
%
\author{Julia A. Meister\and
Raja Naeem Akram \and
Kostantinos Markantonakis}
\authorrunning{J. A. Meister et al.}
\titlerunning{Deep Learning Application in Security and Privacy}
%
\institute{Information Security Group, Smart Card and IoT Centre,\\ Royal Holloway, University of London.\\
\email{julia.a.meister@gmail.com, \{r.n.akram, k.markantonakis\}@rhul.ac.uk}}
\maketitle              
\begin{abstract}
Technology is shaping our lives in a multitude of ways. This is fuelled by a technology infrastructure, both legacy and state of the art, composed of a heterogeneous group of hardware, software, services and organisations. Such infrastructure faces a diverse range of challenges to its operations that include security, privacy, resilience, and quality of services. Among these, cybersecurity and privacy are taking the centre-stage, especially since the General Data Protection Regulation (GDPR) came into effect. Traditional security and privacy techniques are overstretched and adversarial actors have evolved to design exploitation techniques that circumvent protection. With the ever-increasing complexity of technology infrastructure, security and privacy-preservation specialists have started to look for adaptable and flexible protection methods that can evolve (potentially autonomously) as the adversarial actor changes its techniques. For this, Artificial Intelligence (AI), Machine Learning (ML) and Deep Learning (DL) were put forward as saviours. In this paper, we look at the promises of AI, ML, and DL stated in academic and industrial literature and evaluate how realistic they are. We also put forward potential challenges a DL based security and privacy protection technique has to overcome. Finally, we conclude the paper with a discussion on what steps the DL and the security and privacy-preservation community have to take to ensure that DL is not just going to be hype, but an opportunity to build a secure, reliable, and trusted technology infrastructure on which we can rely on for so much in our lives.

\end{abstract}

\section{Introduction}
\label{sec:Introduction}
Computing technology is becoming an integral part of our lives and has many facets ranging from supercomputing (used in weather prediction, cutting-edge research and business automation) to embedded devices (like smartphones, electronic devices in a home and intelligent transport systems). Among many, security and privacy are considered to be two distinct and unique challenges. In the security and privacy domain, any protection system has to match  a constantly evolving adversarial actor. According to the Symantec cybercrime report \cite{symmeticAttackReport}, the overall number of vulnerabilities has increased by 13\% in 2018. Similarly, according to Cybersecurity Ventures \cite{cybersecurityVentures}, zero-day exploits seen in the wild will grow from one per week (in 2015) to one per day by 2021. It is practically impossible for a human to keep pace with the sheer number of cybersecurity events (and related activities) on a daily basis on top of an already daunting threat landscape \cite{ECG-Microsoft}.

In this paper, and as a matter of fact in any context, security and privacy are relative terms. It is not discussed as an absolute state, but rather as a state with potential and/ or accepted risks. The global cost of data breaches has increased by 6.4\% \cite{CostDataBreach} and has the potential to severely damage an organisation's bottom-line, and that is without taking the potential penalties imposed by the General Data Protection Regulation (GDPR) into account \cite{eu:gdpr}. As per the GDPR, an organisation can be fined up to \euro10 million or two percent of the firm's global turnover for a small offence (whichever is greater). For a serious offence, an organisation can be fined up to \euro20 million or four percent of a firm's global turnover (whichever is greater) \cite{eu:gdpr}.

Furthermore, there is a crisis of skilled cybersecurity practitioners. According to study \cite{cybersecurityThreatOpportunities}, the cybersecurity job market will grow by approximately 6 million USD globally by 2019 -- with potential shortages of trained professionals up to 25\%. Automation of decisions and actions based on network and system generated alerts has the potential to help overcome the challenges related to security and privacy -- both in a technological and a business-operations (e.g. labour shortages) dimension. 

Artificial Intelligence (AI) is seen as a potential solution towards the cybersecurity automation challenge in some academic and industrial circles. Machine Learning (ML) has been successfully deployed in a number of domains including but not limited to: image classification \cite{krizhevsky2012imagenet}, objective detection and recognition \cite{ren2015faster}, language translation, and voice synthesis \cite{xiong2018microsoft}. Deep Learning (DL), a type of Machine Learning (ML) method, in most cases does not require prior expert knowledge for its learning (an obvious exception is Neuro-Fuzzy techniques). Therefore, it needs less manually engineered feature extraction and specialist knowledge \cite{deeplearning}. DL can detect patterns in the raw data with potentially higher and more abstract level representations - a function that is very interesting for cybersecurity zero-day vulnerability/ exploit detection. Similarly, DL is used to abstract malware's behavioural features and anomalous activity and can then be used to detect its existence in a system \cite{yuan2014droid,saxe2015deep}. 

AI as a cybersecurity tool is expected to capture a large market and it is clear that AI has the potential to impact the cybersecurity space \cite{Statista}. Furthermore, there is sufficient market interest in both commercial (financial incentives) and academic research. It is understood that there is a potential to mislead an ML/ DL deployment as discussed in existing literature \cite{carlini2017towards,carlini2016hidden}, which is not the focus of this paper. In this paper, we discuss the challenges of deploying AI-based techniques (ML/ DL) to security domains. The discussion highlights the difference between the theory and practice of applying DL methods as a general security tool. The discussed challenges come from the technical development and exploration of DL methods in the context of cybersecurity -- showcasing the fact DL techniques in themselves are not the panacea but mearly a tool that requires a number of correct (and in some cases trustworthy) features to be effective. The robustness of DL is stated in \cite{carlini2017towards} as inversely proportional to the potential of an attacker's ability to find adversarial examples, which can impact the accurate classification and detection of a threat. 

However, in this paper, we argue that robustness, no doubt an important feature, is not just dependent on the attacker's ability to find adversarial examples. It also depends on an interdependent relationship of input data, its accuracy and trustworthiness, potential for adversarial examples, feature richness (needed for accurate classification and detection), and the data representing all possible case scenarios. We will discuss these features in further detail throughout this paper. Furthermore, this paper examines the ML/ DL not only from theoretical and feature/ability specific limitation but also from practical challenges related to implementation and deployment. Existing papers either focus on how successful ML/ DL were in their deployment or specification implementation challenges they faced, but discussion on the challenges related to ML/ DL deployed as security and privacy mechanism are not collectively discusses.  

\subsection{Structure of the Paper}
\label{sec:StructureOfthePaper}
Section \ref{sec:SecurityAndPrivacyViaDeepLearning} elaborates on the existing academic work that has shown the promise of ML/ DL as an automation tool for security and privacy practices. In section \ref{sec:DeepLearningADeeperLookAtItsApplication}, we dive into the technical discussion of DL and how automation based on it is designed and developed. The discussion is derived from first impressions, based on the authors' practical experience coming from a security background. Section \ref{sec:PracticalConsiderationsOfDeepLearningDeployment} articulates the practical considerations that a security practitioner has to take into account when working on DL deployment. Section \ref{sec:ResearchChallengesForDeepLearning} is a list of DL features that would make the technology a useful security tool for cybersecurity practitioners.

\section{Security and Privacy by Deep Learning}
\label{sec:SecurityAndPrivacyViaDeepLearning}
In this section, we survey the types of security and privacy services and applications in which DL is deployed successfully -- as represented by academic literature. 

\subsection{Deep Learning for Security and Privacy}
\label{sec:ProposedApplicationsForSecurity}
The set of security and privacy services that are being explored in academic literature to be the target deployment scenarios for DL are:

\begin{enumerate}
    \item Malware Detection: Efficient pattern recognition in large datasets is what ML/ DL is purpose built for. A number of proposals are put forward in academic literature to identify malware with high accuracy \cite{Yuan:2014, Bojan:2016}. In most of these proposals, pattern recognition is based on a particular behaviour (communication, syscall and resource usage/ utilisation patterns, etc). For an adversarial entity, the objective is to hide or exhibit its behaviour within the scope of genuine applications to avoid detection. 
    \item Anomaly Detection or Network Intrusion Detection: Both the anomaly and the network intrusion detection rely on network traffic analysis. Based on this analysis, ML/ DL techniques try to find usage and communication patterns that represent an abnormal behaviour. It is important to keep in mind that anomalous behaviour is not necessarily a set of activities that are not allowed by system  policies (security/ privacy). It is just an out-of-the-ordinary activity that can be genuine or malicious. For example, user A has access to client records. Usually, user A only accesses one record a day, but today user A accesses the entire list of clients. If the access control policy only focused on access (may user A access client records?) and not on frequency (how many client records user A can access?), accessing all client records would be a permitted action and not suspicious. However, this action might be anomalous. Such classification and detection of out-of-pattern usages nicely fits within the current capabilities of ML/ DL technology \cite{erfani2016high, javaid2016deep, sommer2010outside}.  
    \item Distributed Denial of Service (DDoS) Detection: DDoS can be viewed as an anomalous request to access a particular resource. Therefore, based on the access patterns to a particular resource (a website or an application), ML/ DL can efficiently identify out-of-pattern access requests \cite{zolotukhin2016increasing, yuan2017deepdefense}.  
\end{enumerate}

From the above list, we can ascertain that DL is not widely used for privacy-preservation techniques. There is a potential for exposing data on user access patterns based on the user connection graph, especially in the context of data flow analysis. These domains might have unique patterns that can be useful for an effective DL deployment but an academic literature search for applications of DL in these fields did not yield substantial results. Below, we explain some of the identified privacy related services that might be suitable for DL deployment but limited work has been carried out in academic literature:

\begin{enumerate}
    \item Data Flow Analysis: The flow of data between any two entities can reveal data consumption in an organisation. For example, the flow of data between the consumer database and marketing teams can represent potential value for consumer profiling, targeted marketing and campaign analysis. The data flow and usage in a specific enterprise have a set pattern, even when looking at individual features such as `data flow' and the actual `contents of the data'. Therefore, ML/ DL can be used to identify anomalous usage of data based on the data flows. Anomalous data flow patterns are used by ML/ DL deployed mostly for Intrusion Detection System (IDS) or Intrusion Detection Prevention (IDP) but not as a privacy preservation function. 
    \item Data Exposure Potential: Whether in an enterprise environment or in personal settings, individuals have a circle of other individuals with whom they communicate. A community map for each individual can be constructed based on these communication patterns which can represent not only `with whom' individuals share information but also `what information' is being shared with their community. For example, an individual shares one type of information with only a subset of the individuals in his/ her community. This is easily classifiable and based on the patterns, ML/ DL can predict whether information accessible to an individual at a particular point in time has a high probability of being shared with certain other individuals. This analysis can be used to build a data exposure prediction which can be a useful tool for privacy-preservation and assessment. Furthermore, in the event of an information leakage, an analysis of the data flows and the probability of data exposure can be incorporated into the forensic investigation to quickly find any potential points (individuals) that could have leaked the information. The potential of ML/ DL has not been explored in the context of data exposure in current academic literature. We believe that the application of ML/ DL for such analysis shows a lot of promise. 
\end{enumerate}

Most of the existing literature about privacy and DL is focuses on how to design DL methods in a manner that does not violate the users' privacy \cite{abadi2016deep,li2017multi,shokri2015privacy}.  Another application of DL in privacy is to build recommendation systems for users. For example, Yu et. al. \cite{Fyu2017iprivacy} put forward a privacy setting recommendation system (iPhoto) for photo sharing based on image analysis.  Most dimensions related to DL and privacy are beyond the scope of the this paper. The scope of the paper is how DL itself can be used as a privacy-protection mechanism.

\section{Deep Learning - A Deeper Look at its Application}
\label{sec:DeepLearningADeeperLookAtItsApplication}

In this section, we explore the technical aspects of understanding and deploying DL. The discussion revolves around the pre-requisites for DL deployment, the tools that can be used, and DL optimisation. Readers are referred to consult the survey by Zubair et. al. \cite{fadlullah2017state} for an in-depth analysis of DL structures and methodologies.

\subsection{Representation Learning}
\label{sec:RepresentationLearning}

DL uses representation learning algorithms to automatically identify complex hidden structures in large datasets \cite{deeplearning}. Relations between parameters can be more or less hidden depending on the features present in the data. Representation learning works to solve this problem by transforming raw data into a more useful representation for detection and classification predictors by highlighting the important dependencies \cite{representationLearning}. The challenge is to generalise as much as possible while also preserving most of the information in the original dataset.

DL implements the learning technique in the form of a model, a concatenation of multiple, relatively simple layers that each perform a transformation on the data \cite{representationLearning}. The layers' input is either raw data (input layer) or the previous layer's learned representation of its input (hidden and output layers). This leads to automatically identified, hierarchical levels of abstraction, also called feature extraction, with higher level features defined as a composition of lower-level features \cite{lookingForward, transferLearning}. During the training phase, the model adjusts the internal parameters used to transform the data to achieve a more useful result \cite{deeplearning}.

\subsection{Data Normalisation}
\label{sec:DataNormalisation}

DL models rely heavily on data as it is the basis of the pre-training and training phases, which in turn underlie the specialisation of a model to a task.

DL does not need a perfectly curated dataset due to its learning scheme. Semi-supervised techniques have been shown to alleviate problems, however, a new training strategy and a better cost function could make training on incomplete and noisy data sets more efficient \cite{bigData}. Whitening data is a known way of speeding up training convergence, readers are referred to \cite{efficientBackprop} for details on how to transform the input data.

Ioffe and Szegedy \cite{batchNormalization} describes batch normalisation, where normalisation is embedded in the model architecture as another method to reduce training-times. It works towards fixing the distribution of the layer's inputs and thereby solves the problems introduced by internal covariate shift. Internal covariate shift describes the fact that the layers' input distribution continuously changes during training due to the internal parameters updating \cite{batchNormalization}. The difficulty in changing the dataset in any way is to preserve as much of the original information as possible. This can be achieved by normalising the training examples relative to the entire training data \cite{batchNormalization}. Other, less efficient ways of combating covariate shift include lowering the training rate and careful parameter initialisation.
Using DL in combination with Big Data is a popular concept in the industry, however, there are many challenges that need to be overcome. The three V's model identifies them as volume, variety, and velocity.

Chen and Lin \cite{bigData} provides the authors' thoughts on how to solve these problems. According to the authors, the large volume of Big Data (number of inputs, number of represented classes and high dimensionality of the entries) cannot be accommodated by a single machine due to its limited memory and computing capacity. A distributed framework would be more suited to the task. 
DL has been successfully utilised for the integration of heterogeneous data, e.g. \cite{multimodelDBM} and \cite{multimodelDL}. Therefore, the authors believe that DL methods can be applied to Big Data's large variety of data structures with further optimisation work. 
They propose online learning to combat the velocity (how quickly data is generated).

There are many large data sets ranging across a wide selection of categories publicly available which can be used in training and testing networks. Examples are the MNIST database\footnote{\url{http://yann.lecun.com/exdb/mnist}} of handwritten digits and the Google Audioset\footnote{\url{https://research.google.com/audioset}}, which includes thousands of labelled audio clips. Kaggle\footnote{\url{https://www.kaggle.com}} is a platform that hosts ML competitions and maintains public datasets.
    
\subsection{Designing Deep Learning Models}
\label{sec:DesigningDeepLearningModels}

There are different \textit{neural network architectures} used in DL, each with their own advantages and disadvantages. Convolutional networks are a type of feedforward network that are designed to process multidimensional signals such as images and video \cite{newFrontier}, whereas recurrent networks are adapted to work with sequence data which makes them more difficult to train but applicable to natural language processing (NLP) challenges \cite{empiricalExploration}. Deep Belief Networks (DBNs) are made up of several layers of restricted Boltzmann Machines (RBMs) and are useful for when the training data set is made up of both labelled and unlabelled entries. They often perform better than networks trained only with backpropagation \cite{newFrontier}.

The \textit{training distribution and structure} can be an important factor in the choice of model and learning method. Supervised learning methods require labelled data and tend to have good results when large quantities of data are available \cite{lookingForward}. They adjust the model's internal parameters based on the training loss, calculated by comparing the predicted output to the expected output as defined by the data entry's label. When it comes to unsupervised learning, the ultimate goal is to abstract the raw data in a way that identifies the important factors of variation that apply to all classes. \cite{transferLearning} has had success applying a transductive strategy by using linear models such as Principal Component Analysis (PCA), among others, as some of the network's layers. Semi-supervised learning makes use of both labelled and unlabelled data. The RBMs that make up a DBN are pre-trained with an unsupervised greedy layer-by-layer algorithm and the whole model is then fine-tuned with labelled data and backpropagation. DBNs often perform better than networks trained solely with backpropagation \cite{newFrontier}, as the combination of non-linear layers in a model can be sensitive to the initialisation values. Pre-training, as used with DBNs, can mitigate this sensitivity \cite{lookingForward}.

When it comes to optimising a model's accuracy, tuning the \textit{hyperparameters} is an important step. They are values that directly influence the training of a neural network by configuring a model's complexity and the learning process \cite{hyperparameterOptimization}, both of which are critical to the model's performance. However, finding the ideal values for these parameters can be very difficult as fine-tuning is often based on experience. According to \cite{transferLearning}, there are two common ways of optimising a model's performance through the choice of hyperparameters: manual trial and error and a grid search. Both approaches run into problems when the number of parameters is too large. Readers are referred to \cite{transferLearning} and \cite{randomSearch} for a more efficient optimisation based on random search and greedy exploration. The number and type of parameters differ between models and learning algorithms. Some of the most common include the learning rate, momentum, number of hidden units, number of epochs and batch size.

Training large, distributed networks is slow, as the use of parallel resources is very inefficient. \cite{predictingParameters} introduces a way to reduce the number of free parameters without dropping the accuracy, as many parameters can be predicted and are, therefore, redundant.

\textit{Over- and underfitting} describe situations where a neural network has not learned the ideal generalisation of the training data which leads to poor performance when new data is introduced. This can also be described as the bias/ variance dilemma, a trade-off between high bias and high variance \cite{biasVarianceDilemma}. Common metrics such as training and test error are used to analyse the accuracy of a model can help identify over- and underfitting.

High variance means that a model fails to differentiate between the signal (the general, underlying pattern) and the noise (dataset-specific randomness) of a dataset. In other words, an \textit{overfit model} has failed to sufficiently generalise the features of its specific training distribution and therefore performs poorly on previously unseen data, as it has no general knowledge it can apply. Overfitting can occur with a complex model whose learning algorithm has a low bias and a high variance. Cross validation is a proven method of preventing overfitting by stopping training before the specification becomes to high \cite{earlyStopping}. The point in time at which to stop training is identified by comparing the model's accuracy on the training data to its accuracy on the unseen testing data. Training is stopped if the difference starts growing or is deemed too large, also called early stopping. Reducing the number of parameters is another method of combating overfitting \cite{earlyStopping}. Dropout layers have also been shown to be successful because they prevent the co-adaption of a network's hidden units \cite{lookingForward}. They introduce unpredictable noise into the data by dropping random parameters in each training iteration.

Bias describes the difference between the model's expected output and the correct values. It occurs when the model is oversimplified and does not have enough flexibility to capture the underlying relations of features present in the data or when there are insufficient parameters. A model is said to be \textit{underfit} if it has a low variance but a high bias and can be identified by a high error on both the training and the test data. A possible solution to this problem is changing the model's structure and parameters so that it better fits the problem to be solved.

Bias and variance are inversely related. The ideal model minimises the expected total error of a learning algorithm, which is defined as the sum of squared bias, variance and irreducible error. While bias and variance are reducible, the irreducible error comes from modelling the problem itself.

\subsection{Deploying Deep Learning Methods}
\label{sec:DeployingDeepLearningMethods}

There are many open-source tools and frameworks that support DL which can vary greatly in overhead, running speed and number of pre-made components. Following are short descriptions of a small selection of them.

\textit{TensorFlow}\footnote{\url{https://www.tensorflow.org}} is a Python-based library with automatic differentiation capabilities that support ML and DL. The high-performance numerical computations, modelled as data flow graphs, can be applied to other domains as well. TensorFlow is used by companies such as Google, Uber, and AMD.

\textit{PyTorch}\footnote{\url{https://pytorch.org}} is another such library which enables rapid research on ML networks. The focus lies on extensibility and low overhead, which is possible because the core logic is written in C++. It also supports reverse mode automatic differentiation, which is the most important type of differentiation for DL applications \cite{automaticDifferentiation} and distributed training. In 2017, Uber AI Labs released Pyro\footnote{\url{http://pyro.ai}}, a deep probabilistic programming language (PPL) based on PyTorch.

\textit{Caffe}\footnote{\url{http://caffe.berkeleyvision.org}} is a C++ library that provides interfaces for Python and MATLAB \cite{representationOverview}. It is a clean and modifiable framework, due to the fact that the model's representation is separate from the model's implementation \cite{caffe}. It is very fast in training convolutional networks and allows for seamless switching between devices (CPU and GPU).

\textit{MATLAB}\footnote{https://uk.mathworks.com} can be used for DL among other things and allows users to build and analyse models, even with little expert knowledge in DL. It provides access to models such as GoogLeNet and AlexNet and works with models from Caffe and TensorFlow-Keras. MATLAB also supports collaboration with the PyTorch and MXNet frameworks.

\textit{MXNet}\footnote{\url{https://mxnet.apache.org}} is a very versatile DL framework which supports imperative and symbolic programming as well as multiple languages, such as C++, Python, R, Scala, MATLAB and JavaScript. Its running speed is similar to Caffe and significantly faster than TensorFlow and it is used by both AWS and Azure, among others \cite{representationOverview}.

\section{Practical Considerations of Deep Learning Deployment}
\label{sec:PracticalConsiderationsOfDeepLearningDeployment}
In this section, we discuss the challenges related to deploying DL as part of the cyber security and privacy-preservation mechanism. We discuss three major issues related to the DL, which is in no way an exhaustive list. However, the problems listed in this section have a significant impact on current DL implementations. 

\subsection{Training Data Set}
\label{sec:TrainingDataSet}
Any DL technique requires training to achieve specialisation for a task, therefore the training data set and its structure are very important. There are two crucial elements about the training data set: a) feature-richness and b) trustworthiness. 
Feature-richness means that the training data should be an extensive collection so that the DL model can identify as many features as possible, which will help it identify genuine and malicious behaviours accurately once it is deployed. Features have to be as extensive as possible; for example, data related to an activity should cover as much information about that activity as possible so a malicious entity has as little room as possible to manoeuvre and trick the deployed DL system. Furthermore, the training data should include a diverse set of behaviours. If a training data set is representative of a behaviour set, the algorithm has a better chance of accurately classifying features in it. However, if the behaviour set is not comprehensive, any behaviour that is not part of the set might be miscategorised fand the DL model might not be able to differentiate between a genuine or malicious behaviour correctly.  The reason for this failure is due to the definition or knowledge base for genuine and malicious behaviour is in the behaviour set used for training. Re-enforced learning can be used to accommodate this; however, this can open up a potential avenue for an adversary to modify the behaviour classification of an ML\ DL. 
The second crucial element is the data's trustworthiness. As one of the most important elements of DL, data should be sourced from a trusted environment and this is also true for malicious activities captured (and tagged) in the training data set. The challenge is to capture malicious activities in a trusted manner from a real environment or a lab simulation that accurately depicts how an attacker could behave. Furthermore, the training is carried out on a data set that represents `past' attacks (known attack patterns) and potentially will not be representative of `future' attacks (unknown vulnerability and attack patterns). The challenges related to new and unknown attacks are further discussed in section \ref{sec:ResearchChallengesForDeepLearning}.

\subsection{Adversarial Samples}
\label{sec:PoisonousLearning}
There is extensive work in academic literature that discusses the impact and limitation of ML/ DL against adversarial samples \cite{7467366}. From a deployment point of view, security and privacy practitioners have to keep in mind that a deployed DL system might be susceptible to adversarial samples. This means that an attacker could influence the DL model's training to learn malicious activities as genuine. By doing so, attackers are enabled to accomplish their goal without DL detecting and flagging them. The challenge related to adversarial samples is crucial as organisations deploying DL based security and privacy mechanism would prefer for it to evolve over time, thereby accommodating the increasing sophistication in the threat landscape. However, allowing the evolution of the DL model after initial training opens it up to adversarial samples. On the other hand, if a DL technique is restricted to the initial training then it is not flexible and extensible, two of the important functions DL should have to cope with the challenges of cybersecurity and privacy. A potential middle ground could be to select a DL technique that is the least susceptible and designed to withstand adversarial samples. Unfortunately, even with such methodologies, the likelihood of adversarial samples cannot be completely removed. Therefore, adversarial samples are a threat vector that will see more sophistication in the future as more and more organisations deploy ML/ DL based cybersecurity and privacy-preservation mechanisms. 

\subsection{General Data Protection Regulation (GDPR)}
\label{sec:CollaborationOrIsolation}
Organisations dealing with EU citizens' data have to comply with GDPR regulations. GDPR gives a number of rights to consumers, among which are the two that we are going to discuss in this section: Right-to-Know (RtK) and Right-to-Rectification (RtR). 

Regarding RtK, Article 15.1.h states that ``the existence of automated decision-making, including profiling, referred to in Article 22(1) and (4) and, at least in those cases, meaningful information about the logic involved, as well as the significance and the envisaged consequences of such processing for the data subject'' \cite{eu:gdpr}. This article requires a meaningful information about the processing method used to process their data. As discussed before, DL is chaotic in many instances and the steps taken to reach a particular decision might have limited traceability or support for reverse-engineering. As an example, a user is in his or her rights to  request information on why they received a certain result from an organisation. The organisation then has to explain how the user's data was processed by the company's AI to generate that particular result. GDPR also holds firms accountable for bias and discrimination in their automated decisions. The challenge of explaining how DL has reached a specific decision becomes paramount -- an aspect of the DL that has not been extensively investigated. To what extent DL's choice can be explained and whether that is an acceptable and, more importantly, meaningful explanation to the regulatory-authorities and consumer needs to be further researched. 

RtR (Article 16) states that ``[t]he data subject shall have the right to obtain from the controller without undue delay the rectification of inaccurate personal data concerning him or her'' \cite{eu:gdpr}. If a user exercises RtR, they request changes to their personal data stored in the system. How this change in the data will impact previous processing and leaning, which were then based on incorrect data, is still a big question. The challenge is to make DL rectify its input data selectively post processing in a manner that does not require a complete retraining. 

On a side note, depending on how DL is deployed, the Right-to-Forget, or RtF, (GDPR, Article 17) might have an impact if a sufficient number of consumers/ users request their data to be deleted. At that point, the knowledge set reflecting the behaviour of an organisation's consumers/ users will not be accurate anymore. How this impacts DL's subsequent decisions is still unclear and requires further investigation. 

As a cyberseucrity and privacy practitioner, a clear view of the needs and visions for a DL deployment are necessary. There are plenty of unanswered questions related to DL in terms of research (Section \ref{sec:ResearchChallengesForDeepLearning}), operation, and legislation (GDPR). It is safe to say that this technology has the potential to be beneficial by improving security and privacy-preservation. However, the pertinent question is whether it is ready and mature enough to be deployed extensively as a security and privacy mechanism. The answer to this is complex and depends on multiple factors, including: 

\begin{enumerate}
    \item Organisational requirements and the prioritised security objectives. 
    \item How the organisation envisions using ML/ DL, keeping in mind that ML/ DL are not silver bullets.
    \item Understanding the limitations of ML/ DL and complimenting these techniques with traditional security and privacy measures. 
    \item Accepting that ML/ DL are in the early stage of development and might go through many improvements in the next few years, therefore deployed systems will have to keep up with rapid change (flexibility, extensibility and scalability).
\end{enumerate}

\section{Research Challenges for Deep Learning}
\label{sec:ResearchChallengesForDeepLearning}
In this section, we put forward list of relevant topics and questions for ML/ DL research from the perspective of a cybersecurity practitioner. 

    \begin{enumerate}
        \item Policy change impact analysis: In an enterprise environment, policies change regularly, and can be related to the security and privacy aspects of the enterprise. The impact assessment of such policies on the enterprise environment is based on human experts' knowledge. If the enterprise has deployed ML/ DL as a security and privacy measure, policy changes need to be reflected in the ML/ DL method's learning and execution. To the authors' knowledge, there is no evaluation of how dynamic policies will impact currently deployed ML/ DL implementations. Therefore, predictive impact analysis of policy changes on DL based security and privacy mechanism would be a important step forward. 
        \item Defining a new policy: An organisation's security and privacy objectives are specified by policies and rule-sets. In existing DL, these policies and rule-sets are represented in the labelling of individual records in the training data set. If the policy changes after the deployment of a DL based system, the available option is to generate a new training data set based on the new policies and retraining the DL model. Generating the training data set and retraining can be considered costs in terms of performance and time. The challenge is to cut down this cost and make policy changes as similar to traditional security mechanisms like firewall, access control and IDS, to name a few.  
        \item Preparing DL to cope with the `future': The cybersecurity and privacy landscape is constantly evolving. To cope with this change, DL has to be flexible and have the ability to learn new patterns even after deployment. Furthermore, prior knowledge already learned by a particular instance of DL is valuable, and the ability to transfer it to other instances (for example among multiple organisations) would vastly improve the readiness of the collective cybersecurty field. A potential path forward could be to develop DL techniques with lifelong learning as crucial part.
        \item Isolated or Collaborative Learning: Isolated learning has its pros and cons. The positive side is that as an organisation, your own specific behaviour is profiled. However, this also means that unless you experience a cyber attack, you will not be able to profile it. With collaborative learning, if a single instance of the collaboration experiences a cyber attack, its profiling can then be shared with the other instances in the group. This has the potential of rapidly improving security countermeasures against new and previously unknown attacks. Collaborative learning introduces some additional challenges, such as:
        
        \begin{itemize}
            \item Knowledge based collaboration: In collaborative learning, should algorithms share their knowledge or simply the raw records of the out-of-profile observation? It also requires a method for sharing prior knowledge between multiple DL instances.  
            \item Raw records based collaboration: Sharing raw records seems simple, as each instance can run its own learning process over it. However, this could leak security sensitive data and violate privacy requirements. For raw records based collaboration, efficient and strong anonymisation techniques have to be developed. This anonymisation technique has to protect privacy and security sensitive data but at the same retain sufficient features so that it is still useful for training other DL instances.
        \end{itemize}
        \item Making deep learning forget: There is a number of situations where it is preferable to make the DL de-profile some of the records from its knowledge base. For example, a) the discovery of malicious data in the training data set that is now required to be re-labelled as malicious, b) removing adversarial samples from the DL knowledge and c) if a consumer/ user exercises RtR or right to forget under GDPR. In such situations, DL techniques need to `forget' about certain records. How to achieve this seems to be an open question that will be crucial in a future with increased awareness about privacy in the general public and adversaries successfully training DL implementations with adversarial samples.  
    \end{enumerate}

\section{Conclusion}
\label{sec:Conclusion}
In this paper, we briefly explore the potential, practicality, implications and shortcomings of DL mechanisms in fields such as security and privacy preservation mechanisms. There are numerous proposals in academic literature that advocate the success of DL as an effective mechanism for cybersecurity. We do not evaluate their claims in this paper. We view DL as a mature domain and evaluate how a security practitioner would go about deploying it, what challenges and issues they would have to overcome and what options are available to resolve some of these issues. We do consider that DL has come a long way and can potentially be applied to security and privacy functions with a defined set of static behaviours. In such situations, DL can efficiently detect any behavioural violations with high accuracy. However, it is too early to consider it an extensively useable security measure in its own right. DL has a long way to go before it is mature enough to be deployed as a standalone Unified Threat Management (UTM) environment.
In this paper, we have discussed the aspects of DL an organisation should keep in mind when deploying a DL based solution. In addition, we have also included a list of features that would be useful to security practitioners in number of scenarios if they can be provided by the DL base mechanisms. In conclusion, DL has a lot of promise and with the right features, it could become an impactful tool in the security and privacy arsenal. With the increase of sophistication and complexity of future technology in the current infrastructure, AI-based security and privacy countermeasures (ML/ DL) might be the next logical step. For this reason, cybersecurity researchers have to become active participants in the ML/ DL evolution, rather then just deploying them to security and privacy problems as off-the-shelf kits.

%
%
%
\bibliographystyle{IEEEtran}
\bibliography{paper.bib}

\end{document}